\documentclass[conference]{IEEEtran}
\usepackage[left= 0.625in, right= 0.625in,top=0.76in, bottom=0.76in]{geometry}

\usepackage{cite,graphicx,amsmath,amssymb}
\usepackage{amsfonts,amssymb}
\usepackage{subfigure}
\usepackage{fancyhdr}
\usepackage{mdwmath}
\usepackage{mdwtab}
\usepackage{balance}
\usepackage{xcolor}
\usepackage{bm}
\usepackage{amsthm}
\usepackage{algorithm}
\usepackage{algorithmic}
\usepackage{multirow}
\usepackage{flafter}
\usepackage{mathrsfs}
\usepackage{url}

\theoremstyle{definition}

\newtheorem{corollary}{Corollary}

\renewcommand{\algorithmicrequire}{\textbf{Initialization}}   %
\renewcommand{\algorithmicensure}{\textbf{Optimal algorithm}}

\begin{document}

\title{Resource Allocation for Edge Computing in IoT Networks via Reinforcement Learning }
\author{Xiaolan~Liu, Zhijin~Qin,
~Yue~Gao
\\
Queen Mary University of London, London, U.K.


}

\maketitle

\begin{abstract}
In this paper, we consider resource allocation for edge computing in internet of things (IoT) networks. Specifically, each end device is considered as an agent, which makes its decisions on whether offloading the computation tasks to the edge devices or not. To minimize the long-term weighted sum cost which includes the power consumption and the task execution latency, we consider the channel conditions between the end devices and the gateway, the computation task queue as well as the remaining computation resource of the end devices as the network states. The problem of making a series of decisions at the end devices is modelled as a Markov decision process and solved by the reinforcement learning approach. Therefore, we propose a near optimal task offloading algorithm based on $\epsilon$-greedy Q-learning. Simulations validate the feasibility of our proposed algorithm, which achieves a better trade-off between the power consumption and the task execution latency compared to these of edge computing and local computing modes.
\end{abstract}


\section{Introduction}
The phenomenon of the increasing number of end devices, such as sensors and actuators etc., has caused an exponential growth of requirements for data processing, storage and communications. A cloud platform has been proposed to connect a large number of internet of things (IoT) devices, and a massive amount of data generated by those devices can be offloaded to a cloud server for further processing\cite{fog_survey2018all}. The cloud server generally has an infinite ability of computation and storage, however, it is physically and/or logically far from its clients, implying that offloading big data to the cloud server is inefficient due to intensive bandwidth requirements. Moreover, it cannot satisfy the ultra-low latency requirements for time-sensitive applications and provide location-aware services. Edge computing has been proposed to address this problem by moving data processing to the edge computing devices, such as devices with computing capacity (e.g., desktop PCs, tablets and smart phones), data centers (e.g., IoT gateway) and devices with virtualization capacity, which are closer to end devices, and then a distributed data processing network is implemented\cite{fog_survey2018all,fog2_bittencourt2018internet}. The edge is not located on the IoT devices but as close as one hop to them, or even more than one hop away from them.

Compared with the cloud server, edge devices can support latency-critical services and a variety of IoT applications. The end devices are in general resource-constrained, for instance, the battery capacity and local CPU computation capacity are limited\cite{resource_survey2018resource}. Offloading computation tasks to relatively resource-rich edge devices can meet the quality of service (QoS) requirements of applications as well as augment the capabilities of end devices for running resource-demanding applications\cite{Qos_mach2017mobile}. However, in practice, the computation capacity of edge devices, i.e., a edge server, is finite. Therefore, it cannot support the massive computation tasks from all the end devices in its coverage area. Furthermore, offloading computation tasks of those end devices requires abundant spectrum resources or it might bring about the congestion of wireless channels\cite{Co_mao2017mobile}. Therefore, resource allocation, such as computation capacity, power and spectrum resource allocation, is quite important for such types of resource-constrained networks. Dynamic computation tasks offloading scheme, i.e., the task is executed at a local end device or edge server, is an effective method. It has been mostly discussed in the context of mobile edge computing (MEC), in which the mobile user makes a binary decision to either offload the computation tasks to the edge device or not\cite{binary_bi2018computation}.

Some research work has proposed optimal computation task offloading schemes by minimizing the energy consumption or task execution latency in the network. Most of them have adopted the conventional optimization methods to solve the formulated optimization problem, like Lyapunov optimization and convex optimization techniques\cite{Lyapunov_he2018energy}. However, these optimization techniques can construct an approximately optimal solution only. Note that designing the computation task offloading scheme can be modeled as a Markov decision process (MDP). Reinforcement learning has been adopted as an effective method to solve this optimization problem without requiring the priori knowledge of environment statistics\cite{RL_xu2016online}. But the explosion of the state and action space makes the conventional reinforcement learning algorithm inefficient and even infeasible. Deep reinforcement learning approaches, such as deep Q-network (DQN) has been proposed to explore the optimal policy by solving the aforementioned optimization problem\cite{RL1_chen2018performance,RL_zhang2018deep}.

The increase of computation capacity at edge devices contributes to a new research area, called edge learning, which crosses and revolutionizes two disciplines: wireless communication and machine learning\cite{edge_zhu2018towards,edge1_du2018fast}. Edge learning can be accomplished by leveraging MEC platform. Deep reinforcement learning (DRL) is an effective method to design the computation task offloading policy in wireless powered MEC networks by considering the time-varying channel qualities, harvested energy units and task arrivals\cite{RL1_chen2018performance}. \cite{RL_zhang2018deep} has designed an offloading policy for mobile user to minimize its monetary cost and energy consumption by implementing the DQN-based offloading algorithm. In \cite{ML_wang2018edge}, an "In-Edge artificial intelligence" has been evaluated and could achieve near-optimal performance by investigating the scenarios of edge caching and computation offloading in MEC systems. Moreover, DRL could also achieve a good performance by developing a decentralized resource allocation mechanism for vehicle-to-vehicle communications\cite{V2V_ye2018deep}. Deep learning achieves excellent  performance with large amount of data generated by IoT applications\cite{deep_li2018learning}.

In the previous works, the task execution latency and the power consumption have rarely been considered together when designing the optimal computation task offloading scheme. \cite{MEC_mao2017joint} has optimized the task offloading scheduling by minimizing the weighted sum of execution delay and end device energy consumption with conventional optimization tools. Encouraged by \cite{MEC_mao2017joint}, we formulate a task offloading problem with its objective function including not only the cost in \cite{MEC_mao2017joint}, but also the power consumption of the edge device. Specifically, we are proposing to use reinforcement learning techniques to solve this problem. This approach has been put in use recently to solve the task offloading problem while only considering either the execution delay or the energy consumption as the negative reward function\cite{RL_zhang2018deep,RL1_chen2018performance}. Moreover, we will discuss the remaining computation resource of the end device since it affects the decision making on task offloading when it is run out. The major contribution of this paper is as follows:
\begin{enumerate}
\item We first consider resource allocation in IoT networks with edge computing to design a task offloading scheme for IoT devices. We formulate the weighted sum cost minimization problem with its objective function including the task execution latency and the power consumption of both the edge device and the end device.
\item We solve this optimization problem with reinforcement learning technique. And then we propose the near optimal task offloading algorithm based on $\epsilon$-greedy Q-learning.
\item Numerical results show that our proposed task offloading algorithm achieves a better trade-off between the power consumption and the task execution latency compared to the other two baseline computing modes.
\end{enumerate}

\section{System Model}
\begin{figure}[t]
  \centering
  \includegraphics[width=3.5in]{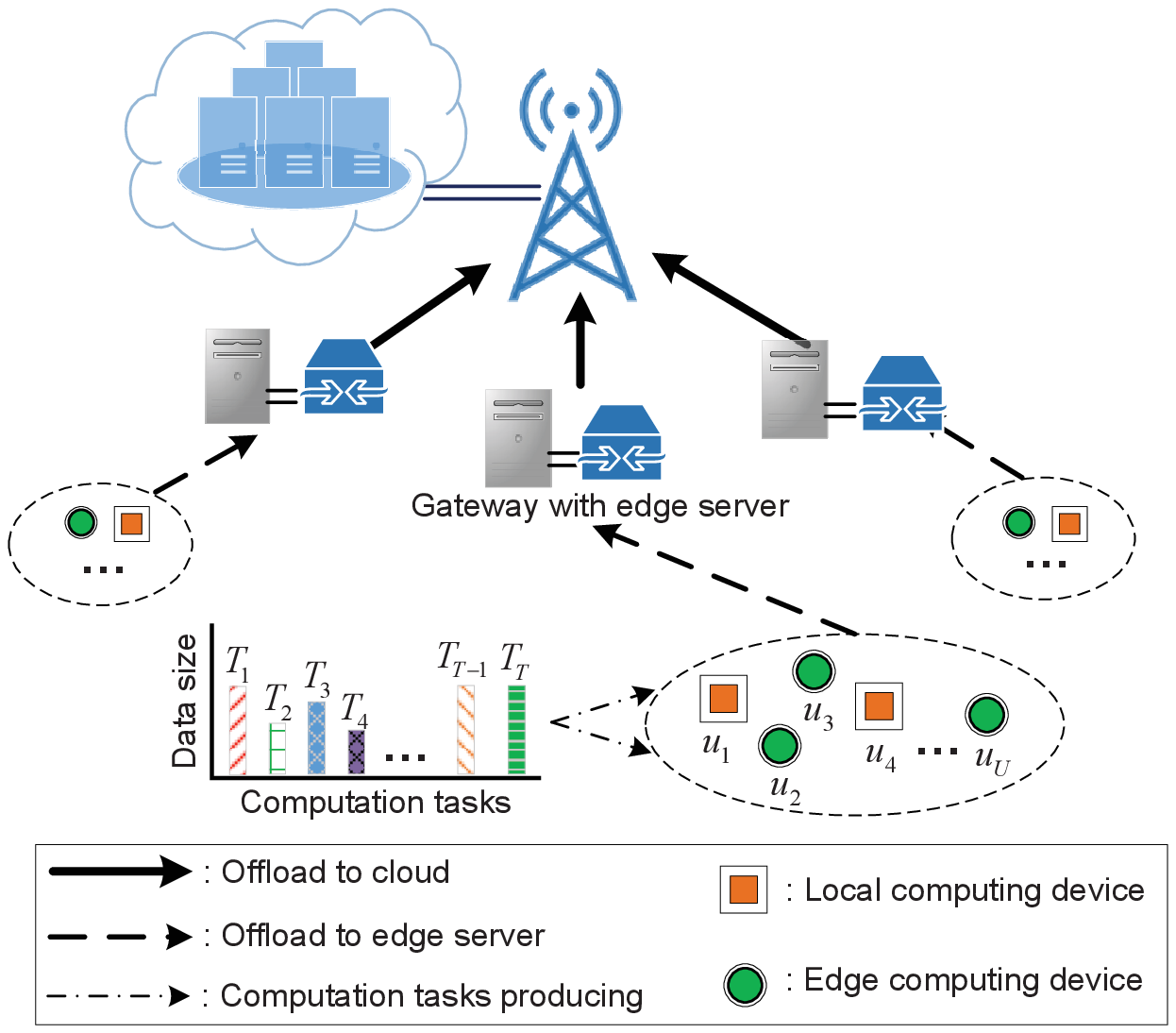}\\
  \caption{Computation tasks offloading model in IoT networks.}
  \label{system model}
\end{figure}

As shown in Fig.~\ref{system model}, we consider an IoT network with many end devices (i.e., IoT devices) and a gateway (i.e., the edge device), where the gateway collects data from end devices in its coverage area and processes them with its equipped edge server. Each end device generates a variety of computation tasks continuously and has limited computation capacity and power, so offloading their tasks to the gateway may improve the computation experience in terms of power consumption and task execution latency.
We focus on a representative end device making its own decisions on task offloading. We discrete the time horizon into epochs, with each epoch equalling to duration $\eta$ and indexed by an integer $0 < k \le K$, $K$ is the maximum number of time epochs in each time horizon. The end device operates over common license-free sub-gigahertz radio frequency and the frequency bandwidth is denoted by $B_w$.
We denote end devices in the network as $\mathcal{U}=\{u_1,...,u_U\}$. The channel condition between the end device and the gateway is assumed to be time-varying. we assume the end device knows some stochastic information about the channel condition in time slot $k$, which is indicated by the channel gain states $\mathcal{G}=\{g_1^k,...,g_G^k\}$ where the channel gain at each time epoch is picking from $G$ possible values. We use a finite-state discrete time Markov chain to model the channel gain state transition over time epochs. Assuming each end device executes a lot of independent computation tasks, these tasks are in different sizes and need to be processed with different CPU cycles. Then we denote the task queue at the end device as $\mathcal{T}=\{T_1,...,T_{max}\}$, where $T_{max}$ is the maximum number of tasks that can be stored at the end device. The task arrival is assumed to be $\mathcal{I}= \{0,1\}$, where $\mathcal{I}= 1$ indicates there is one task generated with its task size randomly picked from $\mathcal{M}=\{m_1,...,m_M\}$, otherwise, there is no task arrived at current time epoch.

From Fig.~\ref{system model}, computation tasks can be either executed at the end device or offloaded to the gateway and executed at the edge server. In each IoT network, some end devices execute the computation tasks locally, while others offload their tasks to the gateway in the same time epoch. At the beginning of each time epoch $k$, each end device makes its own decisions on computation task offloading $\mathcal{O}=\{1\}\cup \{0\}\cup \{-1\}$ and transmit power level $\mathcal{P}_t=\{P_1^k, ...,P_{max}^k\}$ if the end device decides to offload the computation tasks to the edge device. Note that if the end device decides not to offload the computation task when $\mathcal{O}^k=0$, then the cost only contains the local computation power consumption and the local task execution latency, and the transmit power is defined as $P_t^k=0$ in this case. $\mathcal{O}^k=1$ indicates that the end device decides to offload computation task to the gateway, with the transmit power $P_t^k \in \mathcal{P}_t $. In both cases, the computation task is executed successfully, however, if the computation task transmission suffers from outage between end device and the gateway, the computation task execution fails and $\mathcal{O}^k=\{-1\}$.

The task execution latency and power consumption are two critical challenges in edge computing networks, both of them depend on the adopted task offloading scheme and transmit power allocation. In this paper, we consider them as the main cost of our considered IoT network. Therefore, we formulate an optimization problem to minimize the cost function, the weighted sum of the task execution latency and the power consumption.

\subsection{Local Computing Mode}
We have $\mathcal{O}^k=0$ if the computation task is executed at local end device.
We assume the edge server allocates fixed and equal CPU resource for each end device, and it is enough for the computation task to execute in each time frame.
Considering that during any time epoch $k$, $f_d$ denotes the fixed CPU frequency of any end device, which presents the number of CPU cycles required for computing 1-bit of input data. The power consumption per CPU cycle is denoted by $P_d$. Then $f_dP_d$ indicates computing power consumption per bit at the end device. The total power consumption of one computation task at the end device in any time epoch $k$, denoted by $P_{cd}^k$, is given by $P_{cd}^k=f_dP_dm^k$. Moreover, let $D_d$ denote the computation capacity of the end device, which is measured by the number of CPU cycles per second. The remaining CPU resource of the end device in each time epoch is denoted by the remaining percentage of computation resource $\mathcal{R}^k=\{r_{d1}, r_{d2},..., 1\} $. The local computing latency $L_d^k$ is defined as $L_d^k=(f_dm^k)/D_d$. However, the power consumption and the task execution latency are two contradictory challenges in the edge computing network, we cannot reduce them simultaneously, so we are trying to achieve a good trade-off between them. Then we define the cost function of the local computing mode as
\begin{equation}\label{local_cost}
  C_{loc}^k=P_{cd}^k+\beta L_d^k,
\end{equation}
where $\beta$ indicates the weight factor between power consumption and the task execution latency.

\subsection{Offloading Computing Mode}
We assume end devices adopt the time division multiple access (TDMA) scheme to transmit their data to the gateway, that is, the interferences from other end devices are negligible when they are transmitting data over the same time epoch, $k$. Let $g^k$ denote the channel gain from any end device to the gateway, which is constant during the offloading time epoch. $P_t^k$ indicates the transmit power of the end device, then the achievable transmission rate (bit/s) is denoted by
\begin{equation}\label{achievable_rate}
  R_k=B_wlog_2(1+\frac{{P_t^k{g^k}}}{{{\sigma^2}}}),
\end{equation}
where $B_w$ and $\sigma^2$ indicate the bandwidth and the variance of additive white Gaussian noise (AWGN), respectively. Then the power consumption of the end device caused by the data transmission is indicated as $P_t^k$, and the transmission latency is denoted by $L_t^k=T^k/R_k$. Similarly, let $f_s$ denote the computation frequency of the edge server, $P_s$ denote the power consumption per CPU cycle at the edge server. $D_s$ indicates the computation capacity allocated to each end device. The computation power of the edge server is given by $P_{cs}^k=f_sP_sm^k$, and the computation latency is calculated as $L_s^k=(f_sM^k)/D_s$. Therefore, we can obtain the cost function of the offloading computing mode, and it is presented as
\begin{equation}\label{offload_cost}
  C_{off}^k=P_{cs}^k+P_t^k+\beta (L_s^k+L_t^k).
\end{equation}

\section{Proposed Q-learning based Resource Allocation for Edge Computing}
In this section, we formulate our task offloading problem to minimize the weighted sum of power consumption and task execution latency of both the end device and the edge device by optimizing the task offloading decisions, the weight factor and the transmit power of the end device. Since the formulated problem is non-convex, the conventional algorithm is hard or even impossible to solve it, we propose a near optimal task offloading algorithm based on $\epsilon$-greedy Q-learning.

\subsection{Task Offloading Problem Formulation}
Computation tasks from the end device can be offloaded to the gateway depending on the channel conditions, computation task queue and the remaining percentage of the end device's CPU resource. We denote $s^k=(g^k, T^k, r_{d}^k,) \in \mathcal{S}=\mathcal{G} \times \mathcal{T}\times \mathcal{R}_d $ as the network state of any end device in each time epoch $k$. By observing the network state $s^k$ at the beginning of each time epoch $k$, the end device chooses an action $a^k=(O^k, P_t^k)\in \mathcal{A}=\mathcal{O} \times \mathcal{P}_t$ by following a stationary policy $\mathbf{\pi}$. An agent, e.g., each end device, decides whether to offload the computation task and chooses the transmit power level, and we define a penalty function $\delta^k$ as the cost when the task transmission fails. Therefore, the cost function is expressed as
\begin{subequations}\label{cost_function}
\begin{align}
&C^k = C_{loc}^k+C_{off}^k+\delta^k= P_c^k+\beta L^k+\delta^k\\
& \;\;\;\;= P_{cd}^k+P_{cs}^k+P_t^k+\beta(L_s^k+L_t^k +L_d^k)+\delta^k.
\end{align}
\end{subequations}

In this paper, we propose to design an optimal task offloading scheme to minimize the long-term cost of the IoT network, that is, both the immediate cost and the future cost are included. The optimization problem is formulated as
\begin{subequations}\label{optimization_problem}
\begin{align}
\left( \textbf{P1} \right)\;\;
&\mathop {\min }\limits_{\beta ,\;\textbf{P}_\textbf{t},\;{O}} \;\;\;\sum\limits_{k = 1}^K {{C^k}} \\
\text{s.t.}\;\;
&C1:\;0 \le \beta  \le 1;\\
&C2:\;0 \le {{P_t}^k} \le {P_{\max }};\\
&C3:\;{O}^k = \{ 0,1, - 1\}.
\end{align}
\end{subequations}
where $C_1$ denotes the value range of weight factor $\beta$ which balances the power consumption and the task execution latency. $C_2$ is the transmit power of the end device when it decides to offload the computation task to the gateway. $C_3$ presents the task execution set. It is easily noticed that $\textbf{P1}$ is a mixed integer nonlinear programming (MINLP) problem as the integer variable $\mathcal{O}^k$, continuous variable $\textbf{P}_\textbf{t}$ and the discrete variable $\delta^k$ need to be optimized. It is difficult or impossible to find the optimal solution by using conventional optimization techniques. The conventional algorithm has to decouple the optimization problem into many sub-optimization problems and solves them separately, which is inefficient and complicated, so we explore the reinforcement learning techniques to address this problem including multiple optimization variables.

We consider optimizing the variables together in each time epoch, and denote the objective function as a negative reward in (\ref{cost_function}). In addition,
the state transition and cost are stochastic and can be modelled as a Markov decision process, where the state transition probabilities and cost depend only on the environment and the obtained policy. The transition probability $\mathbb P=(s^{k+1}, C^k|s^k,a^k)$ is defined as the transition from state $s^k$ to $s^{k+1}$ with the cost $C^k$ when the action $a^k$ is taken according to the policy. Therefore, the long-term expected cost is given by
\begin{equation}\label{long_term_objective}
V(s, \pi)=\mathbb E_\pi [\sum\limits_{k = 1}^ K \gamma^k C^k ],
\end{equation}
where $s=(g^k, T^k, r_d^k )$, $\gamma \in [0,1]$ is the discount factor and $\mathbb E $ indicates the statistical conditional expectation with transition probability $\mathbb P$.

\subsection{Q-learning Approach}
Generally, the conventional solutions, like policy iteration and value iteration\cite{markov_puterman2014markov}, can be used to solve the MDP optimization problem with a known transition matrix.
But it is hard for the agent to know the prior information of the transition matrix, which is determined by the environment. Therefore, a model-free reinforcement learning approach is proposed to investigate this decision-making problem since the agent cannot make predictions about what the next state and cost will be before it takes each action.

In ($\textbf{P}_1$), each end device is trying to design an optimal task offloading scheme according to some statistical information, such as the possible channel conditions, the possible remaining percentage of computation resource and the possible task queue, observed from the environment. Particularly, we focus on finding the optimal policy $\pi^*$ that minimizes the cost $V(s, \pi)$. For any given network state $s$, the optimal policy $\pi^*$ can be obtained by
\begin{equation}\label{optimal_policy}
 \pi^*=\mathop {\arg \min }\limits_\pi \; V(s, \pi),\; \forall s\in \mathcal{S}.
\end{equation}

The computation task offloading optimization problem at each end device is a classic single-agent finite-horizon MDP with the discounted cost criterion. Then we adopt the classic model-free reinforcement learning approach, Q-learning algorithm, to explore the optimal task offloading policy by minimizing the long-term expected accumulated discounted cost, $C$. We denote the Q-value, $Q(s, a)$, as the expected accumulated discounted cost when taking an action $a^k\in \mathcal{A}$ following a policy $\pi$ for a given state-action pair $(s, a)$.

Thus, we define the action-value function $Q(s, a)$ as
\begin{equation}\label{Q_function}
  Q(s, a)= {\mathbb E_\pi }[{C^{k + 1}} + \gamma {Q_\pi }({s^{k + 1}},\;{a^{k + 1}})|{s^k} = s,{a^k} = a].
\end{equation}
In our proposed algorithm, $Q(s, a)$ indicates the value calculated from cost function (\ref{cost_function}) for any given state $s$ and action $a$, it is stored in the Q-table which is built up to save all the possible accumulative discounted cost. And the Q-value is updated during the time epoch if the new Q-value is smaller than the current Q-value. The $Q(s, a)$ is updated incrementally based on the current cost function $C^k$ and the discounted Q-value $ Q({s^{k + 1}},a),\forall a\in \mathcal{A}$ in the next time epoch.

This is achieved by the one-step Q-update equation
\begin{equation}\label{Q_update_function}
Q({s^k},{a^k}) \leftarrow (1 - \alpha ) \cdot Q({s^k},{a^k}) + \alpha (C^k + \gamma  \cdot \mathop {\min }\limits_a Q({s^{k + 1}},a)),
\end{equation}
where $C^k$ is the cost observed for the current state, $\alpha$ is the learning rate $(0<\alpha \leq 1)$. Q learning is an online action-value function learning with an off-policy, in each time epoch, we calculate the Q-value in the next step with all the possible actions that it can take, then choose the minimum Q-value and record the corresponding action.

Therefore, the computation task offloading optimization problem $\textbf{P1}$ is solved by using the Q-learning algorithm, and to explore the unknown states instead of trusting the learn values of $Q(s,a)$ completely, the $\epsilon$-greedy approach is used in the Q-learning algorithm, where the agent picks a random action with small probability $ \epsilon$, or with $1-\epsilon$ it chooses an action that minimizes the $ Q({s^{k + 1}},a)$ as shown in (\ref{Q_update_function}) in each time epoch.
Then a computation task offloading algorithm based on $\epsilon$-greedy Q-learning is proposed as shown in Algorithm 1.

\begin{algorithm}
\renewcommand{\algorithmicrequire}{\textbf{Initialization}}
\renewcommand{\algorithmicensure}{\textbf{Procedure}}
\caption{Computation Task Offloading Algorithm based on $\epsilon$-greedy Q-Learning}
\label{Q-learning}
\begin{algorithmic}[1]
\REQUIRE~\\
Initialize parameters: discount factor $\gamma$, learning rate $\alpha$, exploration rate $\epsilon$.\\
Initialize action-value function $Q: \mathcal{S}\times \mathcal{A}$ \\
Initialize states: set $g^1$ randomly, set $T^1:=\mathcal{T}$, $r_d^1:=\mathcal{R}_d$. \\
Set $k:=1$,
\ENSURE~\\
\WHILE {$k\leq K \;and \;T >0 \;and \;r_{d}>0$}
\STATE $g^k$ is changed according to a random matrix.
\STATE $e \leftarrow $ random number from [0,1]
\IF {$e <\epsilon$}
\STATE Choose action $a^k$ randomly.
\ELSE
\STATE Choose action $a^k$ according to $\arg \mathop {\min }\limits_{a^k \in \mathcal{A}} Q(s^k,a^k)$
\ENDIF
\STATE Set $s^{k+1}=(G_g^{k+1}, T^{k+1}, r_{d}^{k+1})$, where \\
$ T^{k+1}=T^{k}-a^k+I^k$, \\
$r_{d}^{k+1}=r_{d}^k-a^k(f_dm^k)$.
\STATE calculate the cost $C^k$ by (\ref{cost_function})
\STATE update $Q({s^k},{a^k})$ by (\ref{Q_update_function})
\STATE Set $k:=k+1$
\ENDWHILE
\end{algorithmic}
\end{algorithm}

\subsection{Optimality and Approximation}
The agent in the reinforcement learning algorithm aims to solve sequential decision making problems by learning an optimal policy. In practice, the requirement for Q-learning to obtain the correct convergence is that all the state action pairs $Q(s,a)$ continue to be updated.
Moreover, if we explore the policy infinitely, Q value $Q(s,a)$ has been validated to converge with possibility 1 to $Q^*(s,a)$ , which is given by
\begin{equation}\label{1_possibility}
\mathop {\lim }\limits_{n \to \infty }\mathbb P_r(\left| {Q^*(s,a) - Q(s,a)_n} \right| \ge \varsigma ) = 0,
\end{equation}
where $n$ is the index of the obtained sample, and $Q^*(s,a)$ is the optimal Q value while $Q(s,a)_n$ is one of the obtained samples. Therefore, Q-learning can identify an optimal action selection policy based on infinite exploration time and a partly-random policy for a finite MDP model. In this paper, we approximate the state and action space into finite states, and we use Monte-Carlo simulation to explore the possible policy, so we can obtain a near-optimal policy.

\section{Numerical Results}
\begin{table}
\centering
\caption{Simulation Parameters}
\begin{tabular}{|c|c|}
\hline
 $\gamma$, $\alpha$, $\epsilon$ & 0.5, 0.5, 0.1\\
\hline
 $B_w, \sigma$ & $ 10^5Hz, -174+10log10B_w$\\
\hline
Channel gains & $\mathcal{G}=(0.5, 1, 1.5)*10^{-5}$\\
\hline
Task queue & $ \mathcal{T}=(0,1,...,9)$\\
\hline
Task size &$ \mathcal{M}=(10,11,...,25)Kbits$\\
\hline
$f_s=f_d$, $f_s=f_d$ & $ 500 \;cycles/bit, 10^{-8}\; W \;per \;CPU \;cycles $\\
\hline
$D_s$, $D_d$ & $4GHz, 500MHz $\\
\hline

\end{tabular}
\end{table}

In this section, the performance of our proposed computation task offloading algorithm is verified by the simulations. The task offloading algorithm is applied to each end device so that they can make their own distributed decisions on how to process the generated computation tasks, and this is achieved by using the $\epsilon$-greedy Q-learning algorithm. For comparison, we also provide another two baselines: the local computing and the edge computing mode, which are defined as follows,
\begin{enumerate}
\item  Local computing: the computation task is locally executed at the end device in each time epoch whatever size of the computation task is generated.
\item Edge computing: all the computation tasks generated at the end device are offloaded to the edge device and are processed at the edge server.
\item Proposed scheme: when the task queue is not empty and the end device has remaining computation resource, it decides to execute the task locally or offload it to the gateway at the beginning of each time epoch to achieve the minimum cost, i.e., minimize the weighted sum of cost including power consumption and task execution latency.
\end{enumerate}

\begin{figure}[!t]
  \centering
  \includegraphics[width=3.8in]{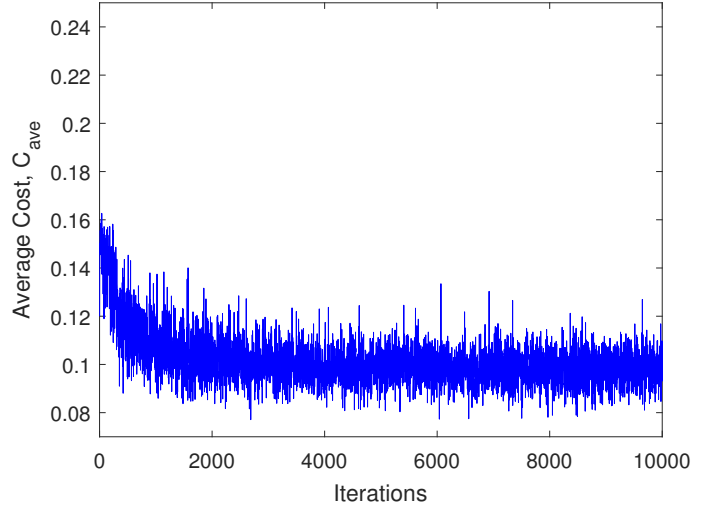}\\
  \caption{Convergence performance of the proposed task offloading algorithm measured by average Cost $C_{ave}$, the weight factor $\beta=0.5$. }\label{convergence_ave}
\end{figure}

\begin{figure}[!t]
  \centering
  \includegraphics[width=3.8in]{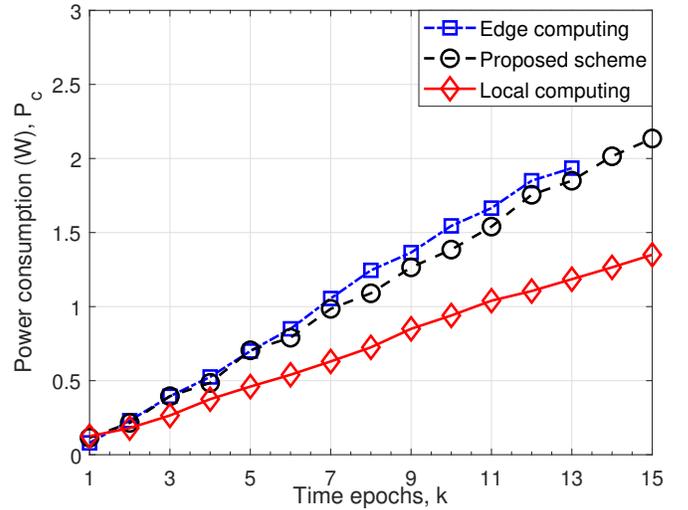}\\
  \caption{Performance comparison of cumulative power consumption $P_c$ with the proposed scheme, edge computing and local computing versus the number of time epochs $k$.  }\label{power}
\end{figure}

We carry out the simulations at any end device in an IoT network. The simulations parameters are displayed in Table I. $\beta=0.5$ is the weight factor of the weighted sum cost, and the time epochs in any time frame is set as $K=15$. Fig.~\ref{convergence_ave} illustrates the convergence performance of the proposed algorithm, where Y-axis presents the average cost, $C_{ave}$, in each time epoch.
We observe that the average cost is converged to a stable value with slight changes.
Based on the convergence of the proposed algorithm, the following numerical results are analyzed.

\begin{figure}[!t]
  \centering
  \includegraphics[width=3.8in]{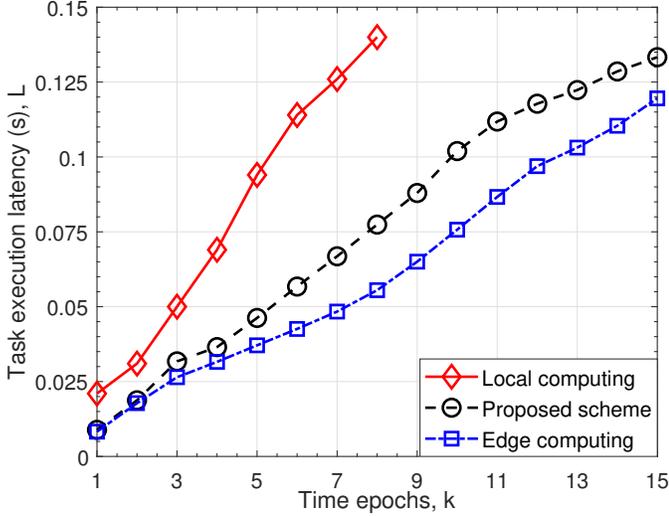}\\
  \caption{Performance comparison of cumulative task execution latency $L$ with the proposed scheme, edge computing and local computing versus the number of time epochs $k$. }\label{Latency}
\end{figure}

\begin{figure}[!t]
  \centering
  \includegraphics[width=3.8in]{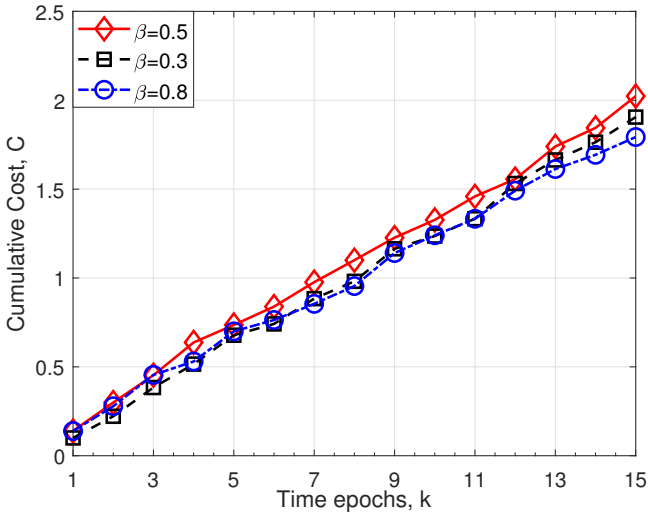}\\
  \caption{Cumulative weight sum of cost $C$ versus different time epochs $k$ with different weight factors $\beta$.}\label{Cost_beta}
\end{figure}

Fig.~\ref{power} shows the comparison of power consumption among edge computing, local computing and our proposed scheme with an increasing number of the time epochs. We can observe that the edge computing has the highest power consumption, this is because the end device consumes transmit power to offload the computation task to the gateway. But for the local computing mode, it consumes the least power since all the power consumption only comes from the task execution. In our proposed scheme, the sequential decisions on computation task offloading are made by continuously observing environment information. The task is only offloaded when it is needed, such that it has less power consumption than the edge computing mode, but has higher power consumption than the local computing mode due to the extra transmit power. Moreover, it is noticed that the edge computing curve finishes earlier than the other two curves. This is because the task execution is stopped and cannot be transferred to the local end device when the task transmission fails.

Similarly, the performance of the task execution latency among the three modes is shown in Fig.~\ref{Latency}. Local computing mode has the highest execution delay since the computation ability of the end device is much weaker than the gateway, and the task size is too large to be processed efficiently with limited compute capability. Correspondingly, the computation task can be executed more quickly by offloading it to the gateway. Our proposed scheme achieves the neutral performance because it makes decisions on the task execution based on the current channel qualities and remaining computation resource of the end device. Specially, from Fig.~\ref{Latency}, we notice that the local computing curve is finished earlier than the other two curves, this is because the limited computation resource of the end device has run out before the end of the time frame.

Fig.~\ref{Cost_beta} illustrates the performance of the cumulative cost with different weight factors $\beta$. It's observed that the worst case happens when $\beta=0.5$ since the cumulative cost, $C$, is higher than others, which implies the task execution latency and the power consumption contribute different weights to the overall cost. It is important to make a trade-off between the two kinds of cost to meet different quality requirements of the computation task.

\section{Conclusions}
In this paper, we investigate the design of a smart computation task offloading policy for the end device in the IoT networks by considering the statistics of environment information, including the time-varying channel conditions, the dynamic task queue and the remaining computation resource of the end device. To solve the formulated task offloading problem, we proposed a $\epsilon$-greedy Q-learning based algorithm to minimize the weighted sum cost of the power consumption and the task execution latency. Our proposed task offloading algorithm has been validated by the numerical results, and the simulations demonstrate that the proposed task offloading scheme achieved a better trade-off performance between power consumption and task execution latency compared to the other two baseline computing modes.

\bibliographystyle{IEEEtran}
\bibliography{IEEEabrv,reinforcement}

\end{document}